\documentstyle[11pt,paspconf,psfig,epsf]{article}

\begin{document}

\title{The X-ray View of AGN -- Overview}
\author{R.M. Mushotzky}
\affil{Code 662, NASA/GSFC, Greenbelt, MD 20771}

\begin{abstract}
Recent {\it ASCA}\ and {\it ROSAT}\ \ X-ray observations of active galaxies have
revealed a host of new data on the fundamental properties of active
galaxies. Amongst these are the discovery and characterization of
absorption by ionized gas in Seyfert-I galaxies (the "warm absorber") ,
the discovery and parameterization of broad Fe K lines which originate
in the central 100 Schwarzschild radii, a substantial modification in the
form of the ionization continuum  from previous models and the
absence of X-ray emission from broad absorption line quasars. We
briefly summarize the present observational situation and indicate
where this field might progress in the next few years with the enhanced
capabilities of {\it AXAF}, {\it XMM} and {\it Astro-E}.
\end{abstract}


\section{Introduction}

X-ray astronomy offers a unique window to observe a wide variety of 
phenomena in active galaxies 
(for a fairly recent, but quite out of date review 
see Mushotzky, Done \& Pounds 1993).  The X-ray continuum is one of the 
main contributors to the total bolometric luminosity , often accounting for 
$> 10$\% of the observed energy.  The X-ray band shows the largest 
amplitude/most-rapid time variability of all the continuum bands indicating 
that it comes 
from the smallest regions. Because of the high penetrating power of X-rays and 
the very broad band (0.1--100 keV) covered by X-ray spectrometers,  X-ray 
spectral data are sensitive to column densities in the range from 
$10^{19}$ to $5\times10^{24}\ {\rm  atms/cm^2}$ (Fig. 1) and the full range 
of ionization states from cold material to 
highly ionized ions like Fe{\sc xxvi} (Figs. 2 \& 4). 
Thus given the relevant spectroscopic 
resolving power, X-ray data allow one to determine total column densities and 
directly measure the ionization fractions. The broad spectral coverage allows 
the observation of a wide variety of physical components representing most of 
the physical regions near the central regions. Many of these components, such 
as the warm absorbers, the power law continuum, the Compton-reflection 
"hump",  the low energy ($E< 0.5$ keV) spectral component connecting the 
X-rays and the UV responsible for ionizing the line emitting clouds, and the 
relativistically-broadened Fe lines, are either not visible or very weak in any 
other spectral band. Thus a detailed understanding of  X-ray spectra are 
necessary for  understanding much of the AGN phenomenon. 
\begin{figure}[t]
\centerline{\hbox{
\psfig{figure=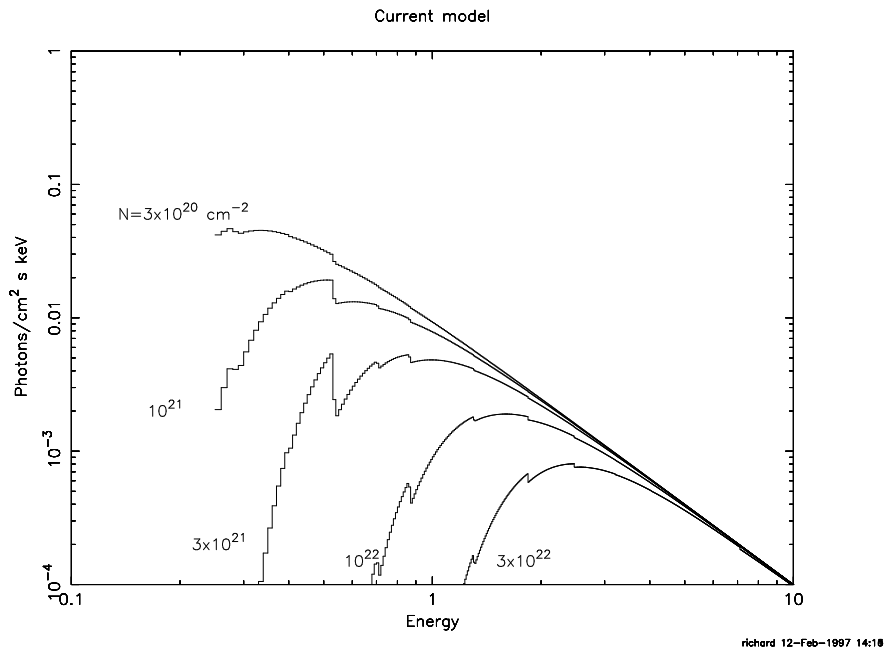,width=12cm,angle=270}
}} 
\caption{The observed photon spectrum of a $\alpha=1$ powerlaw model
absorbed by rest-frame cold material of column density 
0.03, 1,.3, 1 and 3$\times10^{22}\ {\rm atms/cm^2}$. 
Notice at 3$\times10^{22}\ {\rm atms/cm^2}$ a large spectral feature due to 
cold oxygen.}
\end{figure}

\begin{figure}[t]
\centerline{\hbox{
\psfig{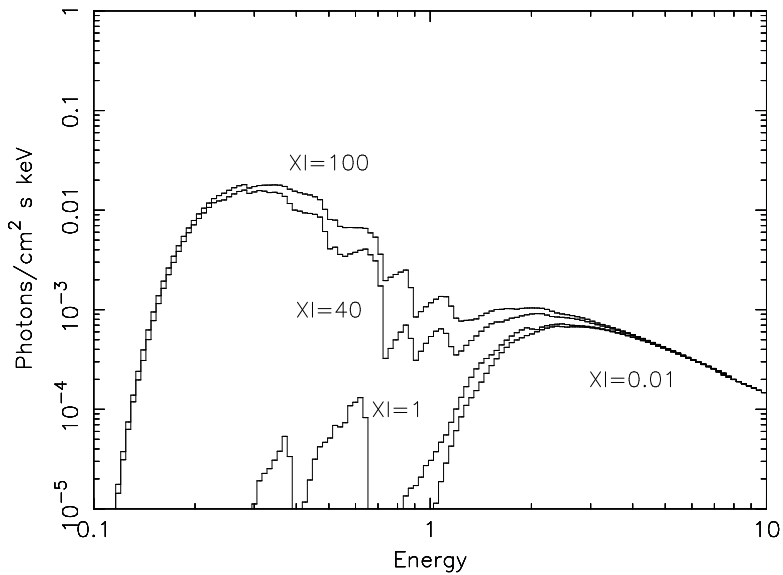}
}} 
\caption{Theoretical models of observed photon spectra, neglecting
resonance absorption features,  of a alpha=1 power law absorbed by a
3$\times10^{22}\ {\rm atms/cm^2}$ absorber at a range of ionization parameters. 
The ionization parameter used is from the {\tt XSTAR} code. 
At low $\xi$  the gas is virtually
unionized and as $\xi$ increase the ionization state rises.}
\end{figure}

The recent large advances in X-ray spectroscopy made possible with 
{\it ASCA}, launched in February 1993,  have revolutionized the field and
point the way to the even larger advances that will be possible with the
launch of {\it AXAF}\ in 1998, {\it XMM}\ in 1999 and {\it Astro-E}\ in 2000. 
In the following
discussion it must be remembered that {\it ASCA}\  has less than $10^{-3}$ of
the Keck collecting area (a milli-Keck) and a spectral resolution of 
$E/\Delta E \sim$10--50.
Despite these limitations {\it ASCA}\  has obtained relatively high quality
spectra for dozens of objects and broad-band spectra for over 150 active
galaxies during its first 4 years of life.

\section{Central Engine}

X-ray emission is a ubiquitous property of active galaxies, even more so than 
broad emission lines or non-thermal optical continuum. As such, it can be 
considered as a fundamental defining characteristic of AGN. Over a fairly 
broad energy range, 0.5--50 keV, the X-ray continuum can be well represented 
by a power law. For broad line objects (e.g. "normal" Seyfert-Is and quasars) 
the X-ray continuum can be approximated by a powerlaw with a narrow range 
of "effective" energy slope 
$\alpha \sim 0.3$--1.2  with a roughly gaussian distribution centered on 
$\alpha \sim 0.7$. 
When the full spectra are well modeled, including the effects of 
Compton-reflection, 
ionized absorption and spectral emission features the mean spectral slope 
changes to $\alpha \sim 0.9$, with a similar dispersion. In Seyfert-I 
galaxies the X-ray 
band often has $\sim30$\% of the bolometric luminosity dropping monotonically 
with luminosity to $<10$\% in the most luminous quasars. However this result 
is based on broad band X-ray fluxes and not on spectral data and is thus 
subject to uncertainties in the spectral bandpass corrections. 

In the small number of Seyfert galaxies with simultaneous X-ray and UV data
there is a strong correlation between the X-ray and UV variability with small
or zero lag, suggesting that a fair fraction (cf. Edelson et al 1996) of the
observed UV radiation is produced by reprocessing of the X-rays. This
possibility, combined with the recent indications (Colina \& Perez-Olea 
1995)  that, at least in radio quiet Seyfert-I galaxies much of the IRAS  IR
luminosity comes from star formation may indicate that the X-ray band is the
only spectral range where the central continuum source is directly observable.

The rapid variability seen in many X-ray sources (e.g. factors of 2 
variability on 200 sec timescales are not uncommon in low luminosity 
Seyfert-I galaxies
(O'Brien \& Leighly 1996; Nandra et al 1997a)), indicates that the X-ray 
continuum originates rather close to the central engine and thus is a prime 
probe of the AGN phenomenon. The rough anti-correlation between X-ray 
luminosity and variance at a fixed timescale 
(Green, McHardy \& Lehto 1993; Nandra et al 1997a) argues for an increase 
in effective source 
size with luminosity. 

\section{Relativistic X-ray Iron K Lines}

The recent discovery (Mushotzky et al 1995; Tanaka et al 1995; 
Nandra et al 1997b) of asymmetric, broad Fe K lines as a common property of 
Seyfert-I galaxies (but not of classical Seyfert-II galaxies, Turner et al 1997)
indicates that the Fe K line is being produced in regions of large
relativistic effects (both special and general). This is, at present, the only
observed spectral feature that originates within $20 R_s$ of the central
engine. There are several Seyfert galaxies (Yaqoob et al 1996; 
Iwasawa et al 1996) where either the flux or the shape of the Fe K line changes
significantly in less than one day, confirming its origin in the very central
regions. A detailed analysis of the largest sample of objects 
(Nandra et al 1997b)
indicates that the gas responsible for the Fe K features is stationary in the
rest frame of the AGN with a net velocity of $< 4000$ km/sec.

\section{No Big Blue Bump}

\begin{figure}[t]
\centerline{\hbox{
\psfig{figure=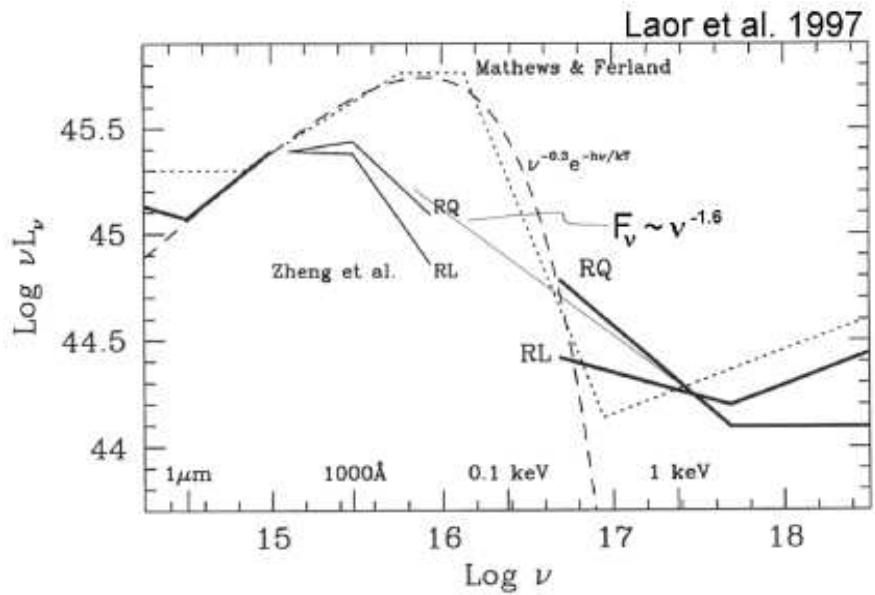,width=12cm}
}}
\caption{The mean UV and X-ray spectra of optically-selected, radio-loud (RL) 
and radio-quite (RQ) quasars as determined by Zheng et al (1997) and 
Laor et al (1997). In the case of RQ quasars, it can be seen that continuum 
in the 13.6--1000~eV can be well modelled by 
${\rm F}_{\nu} \propto \nu^{-1.5}$. This suggests previous 
parameterizations of the ionizing continuum may have overestimated 
the luminosity in this band. As an illustration of this, the 
parameterization of Mathews \& Ferland (1987), and 
form of a 60~eV black-body are shown dotted and dashed respectively.
(Figure adapted from Laor et al 1997.)}
\end{figure}

The combination of high redshift {\it HST}\ observations of quasars 
(Zheng et al 1997), a large {\it ROSAT}\  sample of optically-selected 
quasars (Laor et al 1997)
and a large {\it ROSAT}\  sample of "random" AGN (Walter \& Fink 1993; 
Wang, Brinkmann \& Bergeron 1996) shows that the "Big Blue Bump" as 
parameterized by Mathews \& Ferland (1987) does not exist. 
Instead the continuum over the range of
ionizing radiation (13.6--1000 eV) can be well modeled as a power law of energy
slope $\sim -1.5$ (Fig. 3).
This reduces by factors of 2--4 the total energy available for
radiation driven winds and photoionization of the broad emission lines. The
new continuum in the 10--500~eV band 
does not seem to be a function of luminosity, but 
the ratio of the UV to the hard X-ray decreases as the UV luminosity
increases. 
However that result is based, primarily, on broad-band 
{\it ROSAT}\  and
{\it Einstein} imaging data along with UV photometry, 
and does not take explicit account of
the redshift correction on the form of the continuum. This reduction in the
total luminosity of the EUV band is also important for modeling the ionization
by quasars of the intergalactic medium and the total mass in the 
Ly$\alpha$-forest clouds.

\section{Warm Absorbers}

\subsection{General Properties}

\begin{figure}[t]
\centerline{\hbox{
\psfig{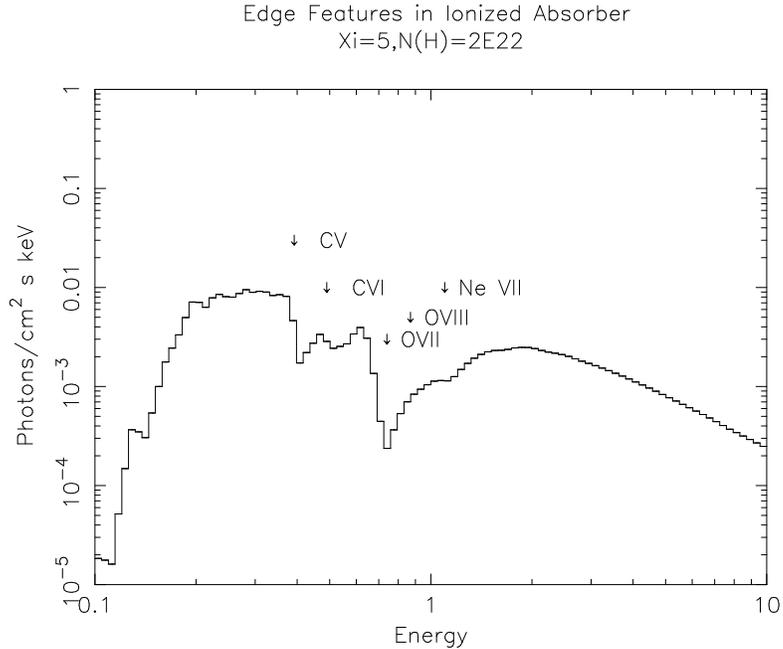}
}} 
\caption{
Theoretical models of  observed photon spectrum, neglecting
resonance absorption features,  of a $\alpha=1$ powerlaw absorbed by a
2$\times10^{22}\ {\rm atms/cm^2}$ absorber at $\xi=5$ indicating the prime 
ions responsible for the edges
in the spectrum. At $E \sim 1.2$~keV there is a shallow  edge due to 
the sum of several Fe L species.
}
\end{figure}
\begin{figure}[t]
\centerline{\hbox{
\psfig{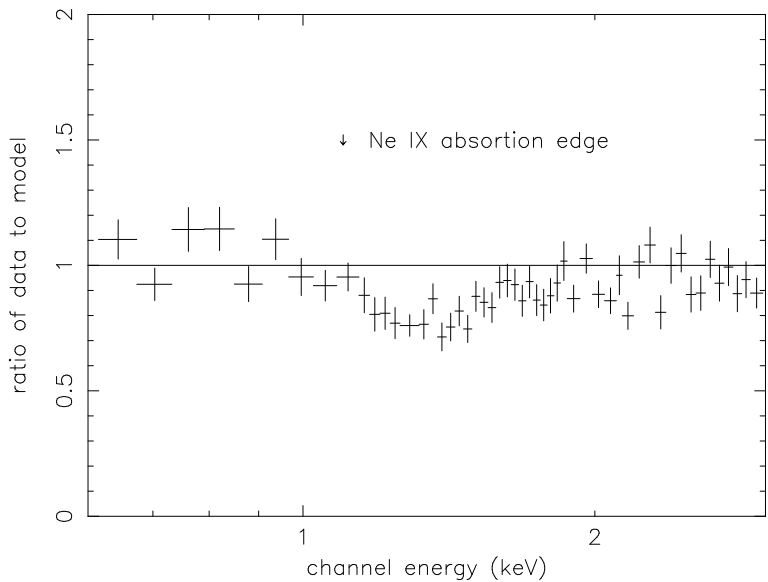}
}} 
\caption{A warm absorber fit to the {\it ASCA}\ spectrum of NGC3783 but
excluding absorption due to neon. The presence of Ne{\sc ix} absorption is
clear.}
\end{figure}

Recent {\it ASCA}\  and {\it ROSAT}\  observations (e.g. Reynolds 1997; 
George et al 1997 for a large sample of objects --- 
see Nandra \& Pounds 1992; Turner et al 1993; Fabian et al 1994
for discovery observations) show that $\sim 1/2$ of bright Seyfert-I
galaxies have absorption features due to the edges of ionized material. Thus
this material has a high covering fraction in Seyfert-I galaxies. The
situation for radio quiet quasars is less clear. 
The {\it ROSAT}\  data (Fiore et al
1994, Laor et al 1997) show that, in general, optically-selected quasars have
optical depths of $< 0.3$ in O{\sc vii} and O{\sc viii}, however this 
limit would not have detected 
$\sim 2/3$ of the O{\sc viii} and 1/3 of the O{\sc vii} Seyfert-I warm
absorbers seen by {\it ASCA}\  (Reynolds 1997). 
Reynolds notes that the optical depths in the {\it ASCA} sample
do seem to drop systematically with luminosity. However there are clearly
luminous objects (e.g. MR2251-179, 3C351, PG~1114+445) which have high optical
depth warm absorbers. As opposed to BALs, there are clearly several radio loud
objects (e.g. 3C212, 3C351) which show warm absorbers. However, 
warm absorbers are not usually seen in the luminous, radio-loud QSOs
(Siebert et al 1996).

At high redshifts the situation is rather unclear because the strongest
features are redshifted into the interstellar absorption band at $z \sim 2$ 
and below the {\it ASCA}\  bandpass at $z \sim 0.6$, thus becoming
very difficult to detect. There are indications from {\it ROSAT}\  hardness 
ratios and spectra for high-$z$, radio-loud quasars that they 
are absorbed (see Section 6 below)
by significant column densities of material, but its ionization state cannot
be determined from {\it ROSAT}\  data.

Because of the relatively low energy resolution of 
{\it ASCA}\  ($E/\Delta E \sim 12$ at 1~keV) edges are more easily detected 
than resonance absorption lines if the
effective velocity width is $< 10,000$ km/sec (but see below for a strong
caveat). Direct fitting of the absorption features are consistent with edges.

The strongest expected edges are those due to the ionized K-shell in C, N, O,
Ne and Fe (e.g. C{\sc v}392, C{\sc vi}490, N{\sc vi}552, N{\sc vii}667, 
O{\sc vi}677, O{\sc vii}739,  O{\sc vii}870 Ne{\sc ix}1100, 
Fe{\sc i}7100--Fe{\sc xxv}9300 eV)  and Fe L-shell transitions in the 
1.0--1.4~keV band.
However N has a relatively low cosmic abundance, the C features
occur at the low energy end of the {\it ASCA}\  bandpass where the collecting 
area is
low and the energy resolution poorer and the signal--to--noise in 
the Fe K band is often low (Fig. 4).
Thus the main features seen in the {\it ASCA}\  data are 
the three O edges, along with Ne K (Fig. 5) and Fe L.

\begin{figure}[t]
\centerline{\hbox{
\psfig{figure=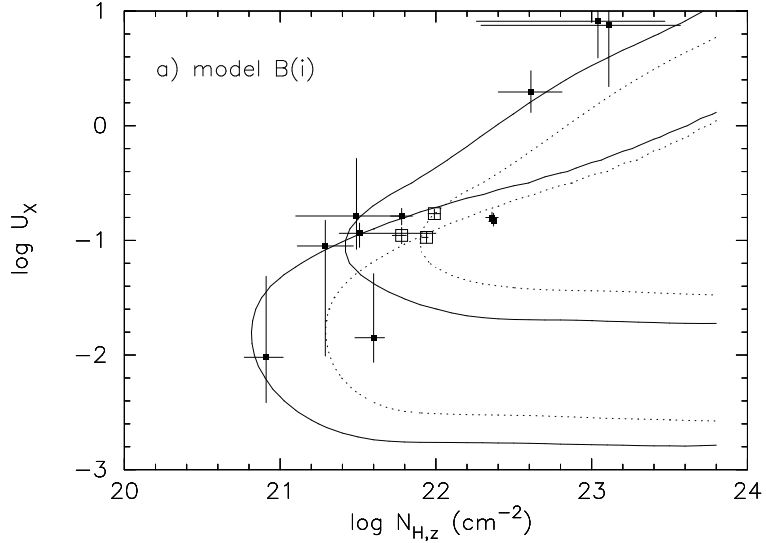,width=10cm,angle=270}
}}
\caption{The column density in photoionized gas ($N_{H,z}$) against the 
'X-ray ionization parameter' ($U_X$, as defined in George et al (1997) over 
the 0.1--10keV band). The filled boxes represent the datasets from an 
heterogeneous collection of Seyfert-Is for which an acceptable fit is 
obtained to the {\it ASCA} data assuming the photoionized gas is in the form 
a uniform screen covering the central source (model {\it B(i)} in 
George et al). The open boxes denote those sources whose {\it ASCA} spectra 
do not satisfy the formal criteria for an acceptable fit in George et al, 
but for which the values of $N_{H,z}$ and $U_X$ are considered reliable.
In both cases the error bars represent the uncertainties on these parameters 
at 68\% confidence. The regions of the $N_{H,z}$,$U_X$ parameter-space in 
which the optical-depth of the O{\sc vii} (lower pair) and O{\sc viii} 
(upper pair) exceeds 0.1 and 2.0 are shown by the solid and dotted curves 
respectively.
(Adapted from George et al (1997).)}
\end{figure}

The 2 strongest features observed are O{\sc vii} and O{\sc viii} 
(Reynolds 1997), while the Ne and Fe L features are only seen in the highest 
signal to noise data.
O{\sc vi} is also sometimes strong
(e.g. NGC~3227, Ptak et al 1994; 
NGC~4593, Kellen et al 1997).
The data can often be well fit by a single
ionization parameter model (George et al 1997) which allows accurate estimates
of the ionization parameter and column densities. George et al find a range of
at least 30 in the ionization parameter and a factor of $>10$ in the fitted
column densities  This corresponds to the locus of points in an ionization
model (Fig. 6) to $\tau$(O{\sc VII}) and/or $\tau$(O{\sc viii}) 
$> 0.1$ and $< 2$. Clearly once
the optical depth is $> 2$ one is not sensitive to the true optical depth and
the limited signal--to--noise ratio in the spectra of most {\it ASCA}\ 
observations obtained to-date 
do not allow features with  optical depth $< 0.1$ to be seen. 
However, it should be noted that there are indications for 
higher ionization, higher column density material 
that can only be detected
by the presence of ionized Fe K edges in 
older, high signal--to--noise ratio {\it Ginga} observations
(Nandra \& Pounds 1994), although these features remain to be 
confirmed.

The combination of high covering fraction and high column density indicates
that this material is the predominant visible component along the 
line--of--sight
to the central object.
Prior to the discovery of the warm absorber, the optical broad-line clouds were 
thought to be the main mass component along the line--of--sight to the 
central engine.
Modelling of these clouds gives column densities in the 
range $10^{22}$--$10^{23}\ {\rm cm^{-2}}$, covering factors 
$<$0.2 (on average) and filling factors of $10^{-3}$. Thus the 
warm absorbers, with a similar column density, larger covering factor and 
somewhat larger (but uncertain) filling factor, must have a larger 
total mass.

\subsection{Velocity}

Direct fits of red/blue shifted models to the {\it ASCA}\  data show no 
evidence for
"motion" with typical upper limits being $v< 10,000$ km/sec. However 
the {\it ASCA}\ 
data are primarily sensitive to the energy of onset and cannot sensibly
constrain the "width" of the edge, due primarily to its large intrinsic width.
Thus this type of analysis would not have found the BAL phenomenon.

\begin{figure}
\centerline{\hbox{
\psfig{figure=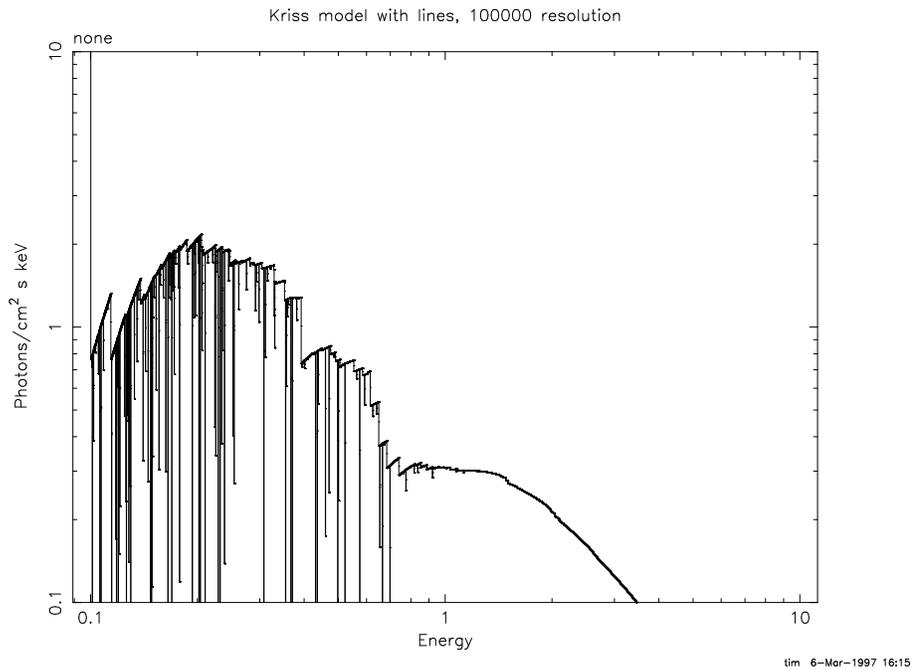,width=12cm,angle=270}
}}
\caption{Theoretical photon spectrum from the {\tt XSTAR} code 
including narrow resonance absorption and edges. This model is similar to 
one used by Kriss \& Krolik (1995). Notice the numerous optically-thick  
resonance
absorption predicted in the model at $E < 1$~keV. This model has 
not been convolved with the {\it ASCA} energy resolution.}
\end{figure}
However, the general absence of strong resonance absorption features, can, in
principle, set limits to the velocity field (Kriss et al 1996). The general
point (Fig. 7) is that in a photoionized plasma there are large numbers of
optically thick resonance absorption lines which can be closely spaced. Thus
if they are sufficiently broadened to overlap in energy they can be detected
even with a low resolution spectrometer. Kriss et al (1996) use such a model to
derive an upper limit to the Doppler-width
$b<200$ km/sec from the {\it ASCA}\  spectra of NGC3516. 
As pointed out by Mathur et al (1996), if the UV and X-ray absorbers are due
to the same material one derives even smaller values for the $b$ parameter.
While these arguments seem to be rather tight, the low resolution 
{\it ASCA}\ data do not allow a direct test. The $R \sim 500$ {\it AXAF}\ 
and {\it XMM}\ spectra should allow a direct measurement of the velocity field 
of the warm absorbers. If the UV and X-ray absorbers are the same, this 
implies a wind with a fairly high mass flow rate
($\sim$1 ${\rm M_{\circ}/yr}$ in MCG-6-30-15)
$M(outflow) \sim R_{pc} N_{22} v_{8} (\Omega/4\pi)\ {\rm M_{\circ}/yr}$, 
where $R_{pc}$ is the radius at which the warm
absorber exists in parsec, $N_{22}$ is the column density in units of 
$10^{22}\ {\rm atms/cm^2}$, $v_8$ is the outflow velocity in units of 
$10^8$ cm/sec, and 
$\Omega/4\pi$ is the covering factor
(Reynolds 1997). Of course, the X-ray data themselves do not 
require a wind solution.

\subsection{Location and Density}

Direct fits of single parameter (e.g. unique values of ionization parameter
and column density) ionized absorber models often provide good fits to the
data (e.g. George et al 1997). However there are examples (Guainazzi et al
1996, George et al 1997) where such models do not describe the data well,
indicating a more complex physical situation. It must be remembered that not
all of the ionized absorber models make the same assumptions regarding
resonance absorption lines, emission lines from the ionized gas or the form of
the ionizing continuum and thus it is not clear at present if the deviations
of the data from the model are due to the neglect of relevant physics or a
more complex physical situation.

In principle one can use the time variability in the observed optical depths,
combined with the equations for the photoionization and recombination time
scales to set limits on the location of the warm absorbers. While such data
exists for several objects (NGC 3227, Ptak et al 1994; 
MCG-6-30-15, Otani et al 1996; NGC 4051, Guainazzi et al 1996; 
NGC4593 Kellen et al 1997),
the detailed analysis have only been done for 
MCG-6- 30-15 and NGC4051.

Using the definition of ionization parameter 
$\xi=L/nR^2=L dR / N R^2$, where $N$
is the column density and $dR$ is the thickness of the slab, then
$dR/R=\xi N R/L$.

The recombination timescale to all levels for highly ionized oxygen is 
$t_{rec} \sim 100T^{0.7}/n_9$ sec, where $n_9 =n/10^{9}$ and 
$n$ is the equivalent hydrogen density . 
This can be expressed as 
$t_{rec} \sim 100\xi^2 R_{16}^2 L_{43}^{-1} T_5^{0.7}$~sec
and the photoionization timescale for a given ion is
$t_{ph} \sim 4x10^{10}/n$~sec 
$\sim 20 R_{16}^{2} L_{43}^{-1}$~sec.

As the continuum increases in intensity one, naively, expects the gas in the
line of sight to become more ionized on a timescale $t_{ph}$ and when the
continuum declines in intensity the gas should recombine and become less
ionized on a timescale $t_{rec}$. The anticipated timescales are amenable to
direct observation.
 
The data for NGC4051 and MCG-6-30-15 do not show the expected behavior, in
that the O{\sc vii} features remain relatively constant with time while the
O{\sc viii} features vary. Thus one is  led to two zone model.  In the outer
zone, responsible for the O{\sc vii}, the gas is low density and the
photoionization and recombination timescales are long and thus the gas does
not respond to the continuum, In the inner region, responsible for O{\sc viii}
the gas is responding rapidly. 
In MCG-6-30-15 the recombination time  results indicate
that O{\sc viii} comes from a region consistent with the 
optical/UV BLR and O{\sc vii}
from further out, and that the O{\sc viii} region has a small filling factor.
The applicability of a two-zone model is called into question by the strong
overlap in ionization parameter space between O{\sc vii} and O{\sc viii} 
(cf. Fig. 4 of Otani et al 1995). Over roughly half of the parameter space 
for which O{\sc vii} is strong one has O{\sc viii} and there is only a 
narrow range of parameter space in which O{\sc vii} is dominant and there is 
little or no O{\sc vi} or O{\sc viii}.
I thus suspect that the two-zone models are somewhat fine tuned.
However, this solution assumes that the gas is in thermal equilibrium which is
not necessary (Kriss \& Krolik 1995). Clearly more spectral features are
necessary to determine the physical state of the gas.

\begin{figure}[t]
\centerline{\hbox{
\psfig{figure=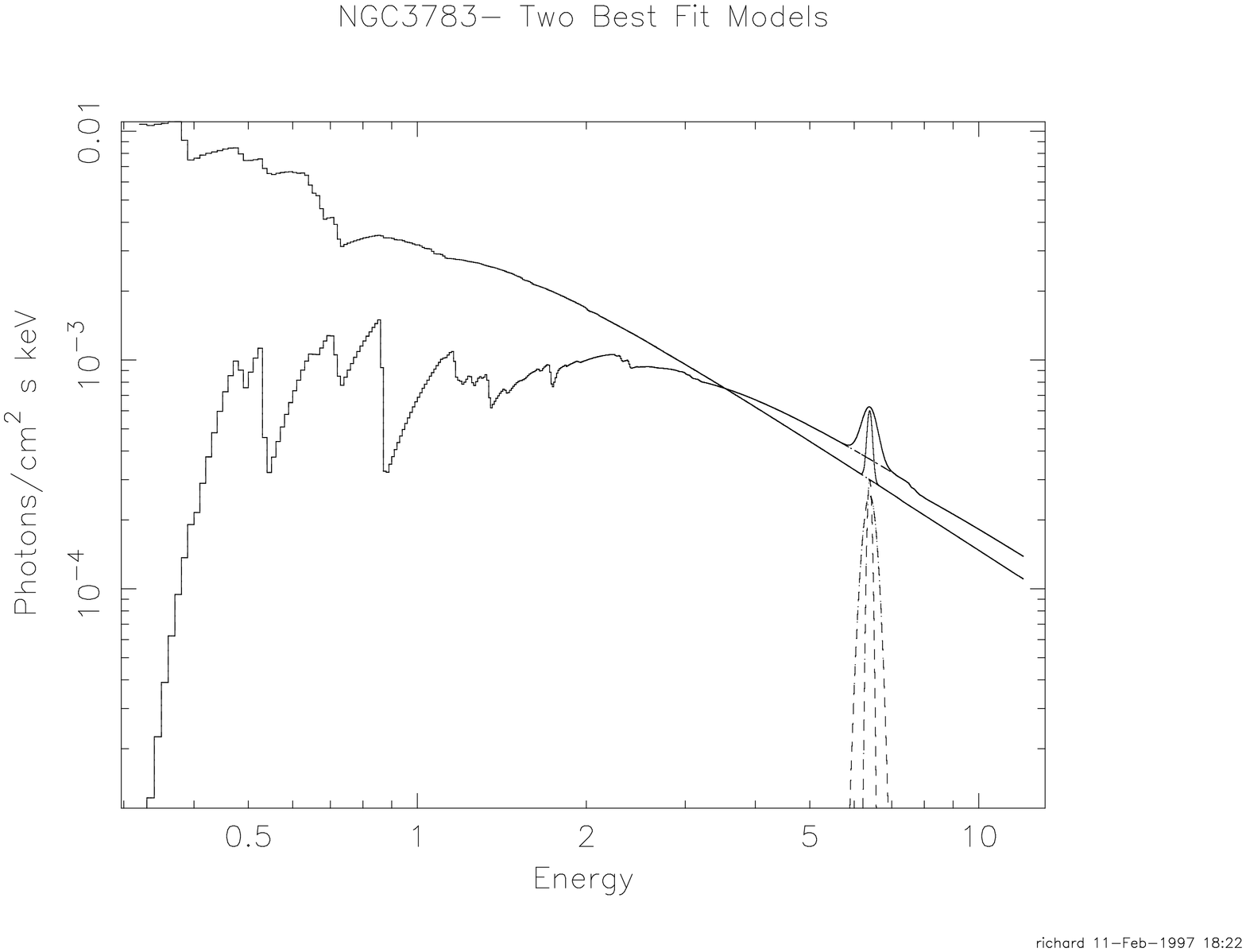,angle=270,height=8cm}
}}
\caption{Comparison of the best fitting warm absorber models to 2 epochs of 
the {\it ASCA}\ NGC3227 data. 
The second epoch data (the lower of the two lines) requires
a ionization parameter at least 10 times larger and a column density 5 times
larger than the first epoch, in addition a larger column  density of cold
material was required. The data were taken 2 years apart in 1993 and 1995.}
\end{figure}
The generality of these results is not yet known, for example NGC3227 
(Ptak et al 1994) seems to behave somewhat differently (Fig. 8).
The depth of the
oxygen features and the nature of the prominent features all changed, as did
the implied column density of low ionization material but the flux in the
power law component remained constant . This data  are best explained as a
change in the material in the line of sight rather than as a ionization change.

Potentially, measurement of the strength of emission compared to absorption
features will allow a measurement of the overall geometry of the system and a
check of ionization equilibrium assumptions (Netzer 1996). A sufficiently well
sampled time series can also allow a reverberation mapping of the warm
absorber region. However the limited energy resolution 
of {\it ASCA}\  makes
it difficult to determine the strength of the relatively weak emission lines
that are predicted.

\subsection{Correlation with other properties}

There is a strong statistical correlation between the presence of UV "narrow"
absorption and warm absorbers (Crenshaw 1997; Mathur 1997): which is examined
in detail in Mathur et al (1996) and is put in the context of a physical wind
model by Murray \& Chiang (1996). However, one must remember that a comparison
of the UV and X-ray absorption features (Kriss et al 1996)  is a strong
function of the continuum model used (Mathur 1997) and the
photoionization code.

While warm absorbers  tend to occur more frequently in low luminosity objects 
so far they have not been detected in any  LINERS. However, the signal to
noise in these observations is often lower than that in the Seyfert-I data.

The presence of warm absorbers in Seyfert-II galaxies is unclear (Turner et al
1997) due to the overall spectral complexity of these objects. However, they
often show large column densities of cold material and at least in a few cases
soft X-ray spectra consistent with scattering from a ionized region whose
ionization parameter and column density are not consistent with that seen in
Seyfert-Is.

\section{BAL Quasars}

In a major surprise, optically-luminous BAL QSOs are very weak X-ray sources
(Green \& Mathur  1996) often being 30--100 times less luminous in X-rays
than expected from their optical luminosities. This is seen in both the large
sample from the {\it ROSAT}\  all-sky survey data and from 
the more sensitive pointed
data.  Thus either BALs are intrinsically X-ray quiet or  
the {\it ROSAT}\  flux is reduced by high column densities 
($N(H) > 5\times10^{22}\ {\rm atms/cm^2}$, Green \& Mathur
1996) of either cold or ionized material. The distinction between these two
possibilities  requires a higher spectral resolution observation.

The only {\it ASCA}\  observation of a classical BAL 
(PHL5200, Mathur, Elvis  \& Singh  1995) seems to confirm the high column 
density, but the data are of relatively poor signal to noise. 
Unfortunately the only other "classical" BAL
observed by {\it ASCA}, PG1416-129, is not a true BAL.

During this meeting several of us have realized (Turnshek 1997; Wills 1997)
that there are several recently discovered BALs with high quality 
{\it ASCA}\  data -- we await the results of the analysis.

The absence of selective reddening in BALs combined with the high inferred
X-ray column densities argues for  very little dust - which may be "natural"
if the gas is highly ionized. Similar results are seen for several IRAS
selected AGN (Brandt, Fabian \& Pounds 1996) with {\it ASCA}\  spectra. 
The existence
of highly ionized gas interspersed with dust is a new  physical regime which
is not yet fully understood.

The general absence of X-ray BALs (e.g.troughs due to broad resonance
absorption of O{\sc vii}, Ne{\sc viii} etc - see Fig. 7) is a 
little hard to understand unless
the ionization parameter is finely tuned, or the line widths are "narrower"
({\it ASCA}\  $E/\Delta E \sim 3$--10,000 km/sec) for higher ionization 
objects. 
While there has
been no systematic search for them it is clear that, at least in low redshift
objects, features of optical depth $>0.2$ which might be associated with 
O{\sc vii} He-like or O{\sc viii} H-like resonance absorption 
troughs, are in general absent.
The recent possible detection of X-ray BALs (Leighly et al 1997) in a set of
narrow line Seyfert-I galaxies may help us to understand this phenomenon.

\section{Conclusions}

The missions to be launched in the next 3 years ({\it AXAF}, {\it XMM} \& 
{\it Astro-E})\footnote{for further information see 
\verb+http://heasarc.gsfc.nasa.gov/docs/heasarc/missions.html+}  
will improve, compared to {\it ASCA}\ , spectral resolution by 10--100, 
collecting area by
factors of 3--10 and bandpass by factors of 2--4, 
similar to the improvements
that {\it ASCA}\  represented over previous experiments. 
The possibility of having
very high signal to noise CCD type spectra combined with $R \sim 100$--1000
high resolution spectra and spectral timing sensitivity for absorption features
down to $\sim 5000$~sec promises to truly revolutionize X-ray spectral
studies of absorption in AGN. The vastly increased power will allow the
measurement of many 10's--100's of objects which will allow a direct connection
to large radio, UV and optical samples.

\acknowledgments
I would like to thank Dr. R. Weymann for the invitation to this exciting
meeting, Dr. T. Kallman for useful discussions and for providing the 
{\tt XSTAR} models, and
Dr. I George for communication of results prior to publication, 
a careful reading of the text, and assistance with its preparation.


\begin{references}
\reference Brandt, W.N., Fabian, A., Pounds, K.,
	1996, \mnras, 278, 326 
\reference Colina, L, Perez-Olea, D.,
	1995, \mnras, 277, 845
\reference Crenshaw, D.M., 
	1997, these proceedings
\reference Edelson, R.  et al., 
	1996, \apj, 470, 364
\reference Fabian, A. et al., 
	1994, \pasj, 45, L59
\reference Fiore, F., Elvis, M., McDowell, J., 
		Siemiginowska, A., Wilkes, B., 
	1993, \apj, 431, 515 
\reference George, I.M.  et al 
	1997, in prep
\reference Green, P., Mathur, S.,
	1996, \apj, 462, 637 
\reference Green, A.R., McHardy, I., Lehto, H.,   
	1993, \mnras, 265, 664 
\reference Guainazzi, M., Mihara,T., Otani, C., Matsuoka, M. 
	1996, \pasj, 48, 781
\reference Iwasawa, K. et al.,  
	1996, \mnras, 282, 1038
\reference Kellen, M., et al 
	1997, in prep
\reference Laor, A., Fiore, F., Elvis, M., Wilkes, B., McDowell, J., 
	1997, \apj, 477, 93 
\reference Leighly, K., Mushotzky, R.F., Nandra, K., Forster, K.,
	1997, \apjl, submitted
\reference Kriss, G. et al.,
	1996, \apj, 467, 629 
\reference Mathur, S., 
	1997, these proceedings
\reference Mathur, S., Elvis, M., Wilkes, B., 
	1995, \apj, 452, 230
\reference Mathur, S., Elvis, M., Singh K.P.,   
	1995, \apj, 455, L9 
\reference Mathur, S., Wilkes, B., Aldcroft, T.,
	1997, \apj, 478, 182
\reference Mathews, W., Ferland, G.,
	1987 \apj, 323, 456
\reference Murray, N., Chiang, J.,
	1996, \apj, 454, L105 
\reference Mushotzky, R., Done, C., Pounds, K., 
	1993 \araa, 31, 717
\reference Mushotzky, R. et al. 
	1995, \mnras, 272, L9 
\reference Nandra, K., Pounds, K.,
	1992, Nature, 359, 215 
\reference Nandra, K., Pounds, K.,
	1994, \mnras, 268, 405
\reference Nandra, K., George, I.M., Mushotzky, R., Turner, T.J., Yaqoob,T., 
	1997a, \apj, 477, 70  
\reference Nandra, K., George, I.M., Mushotzky, R., Turner, T.J., Yaqoob,T., 
	1997b \apj, 477, 702 
\reference Netzer, H., 
	1996, \apj, 473, 781 
\reference O'Brien, P., Leighly, K., 
	1996, Adv Space Res, in press (astro-ph/9701105)
\reference Otani, C. et al., 
	1996, \pasj, 48, 211
\reference Ptak, A., Yaqoob,T., Serlemitsos,P.J., Mushotzky, R., 
	1994, \apj, 436, L31 
\reference Reynolds,C.  
	1997, \mnras, in press 
\reference Siebert, J., Matsuoka, M., Brinkmann,W., Cappi,M., 
	Mihara, T., Takahashi, T., 
	1996, \aap, 307, 8 
\reference Tanaka,Y. et al 
	1995, Nature, 375, 659
\reference Turner, T.J., Nandra, K., George, I.M., Fabian, A., Pounds, K.,  
	1993, \apj, 419, 127 
\reference Turner, T.J. et al 
	1997, \apjs, submitted  
\reference Walter, R., Fink, H. 
	1993, \aap, 274, 105 
\reference Wang, T., Brinkmann, W., Bergeron, J.,
	1996, \aap, 309, 81   
\reference Yaqoob, T., Serlemitsos, P.J., Tuner, T.J., George, I.M., 
	Nandra, K.,
	1996, \apj, 470, L27 
\reference Zheng, W., Kriss, G., Telfer, R., Grimes, J., Davidesen, A., 
	1997, \apj, 475, 469 
\end{references}
\end{document}